\documentclass[prd,twocolumn,showpacs,superscriptaddress,preprintnumbers,epsf]{revtex4}

\usepackage{graphicx}
\usepackage{dcolumn}
\usepackage{bm}
\usepackage{epsf}

\renewcommand{\bar}[1]{\overline{#1}}

\usepackage{amssymb}

\renewcommand{\bar}[1]{\overline{#1}}

\usepackage{indentfirst}
\usepackage{psfig,color}
\usepackage{epsfig}
\usepackage{epsf}
\usepackage{graphicx}

\providecommand{\Journal}[4] {#1 {\bf #2}, #3 (#4)}
\providecommand{\EPJA}{Eur. Phys. J. A } %
\providecommand{\MPLA}{Mod. Phys. Lett. A} %
\providecommand{\NPA}{Nucl. Phys. A } %
\providecommand{\NPB}{Nucl. Phys. B } %
\providecommand{\PLB}{Phys. Lett. B } %
\providecommand{\PRL}{Phys. Rev. Lett. } %
\providecommand{\PRC}{Phys. Rev. C } %
\providecommand{\PRD}{Phys. Rev. D } %
\providecommand{\PRSA}{Proc. Roy. Soc. A } %
\providecommand{\ZPA}{Z. Phys. A } %

\preprint{hep-ph/0311331}

\begin{document}


\title{Parity of Anti-Decuplet Baryons Revisited from Chiral Soliton Models}

\author{Bin Wu}
\affiliation{Department of Physics, Peking University, Beijing
100871, China}
\author{Bo-Qiang Ma}
\email{mabq@phy.pku.edu.cn} \altaffiliation{corresponding author.}
\affiliation{ CCAST (World Laboratory), P.O.~Box 8730, Beijing 100080, China\\
Department of Physics, Peking University, Beijing 100871, China}

\begin{abstract}
We recalculate masses and widths of anti-decuplet baryons in the
case of positive parity from chiral soliton models, provided that
the member $\Xi_{3/2}$ of the anti-decuplet has a mass 1.86~GeV as
reported recently. Calculations show that there are no convincing
candidates for the nonexotic members of the anti-decuplet
available in the baryon listings. Up to the leading order of $m_s$
and 1/$N_c$, the width formula for the decay of the anti-decuplet
baryons to the octet depends only on $SU(3)$ symmetry
model-independently, except the coupling constant. Similarly we
give a width formula for the decay of negative parity baryons
belong to certain $SU(3)$ baryon multiplet by pure symmetry
consideration. By this formula, we find that if we have an
anti-decuplet with negative parity and that the masses are the
same as those given by chiral soltion models, the identification
of $N(1650)$ as $N_{\bar{10}}$ are inconsistent with experiments
for $N(1650)\rightarrow N\pi$ while the widths agree with other
two decay channels involving strangeness. And $\Sigma(1750)$ seems
to be a reasonable candidate for $\Sigma_{\bar{10}}$.
\end{abstract}


\vfill

\vfill

\maketitle

Due to recent reports \cite{LEPS,DIAN,CLAS,SAPH,HERMES} for the
existence of pentaquark state $\Theta^+$, Skyrme's old idea
\cite{Skyr} of identifying baryons as solitons has aroused
unprecedent interest. In the large $N_c$ limit \cite{witt}, the
soliton picture of baryons can be proved consistent with QCD. The
quantization of the SU(3) Skyrmion in collective coordinates
\cite{Guad} not only gives the correct baryon octet and decuplet,
but also predicts the next lighter baryon multiplet, the
anti-decuplet \cite{Mano-Chem}, in which $\Theta^+$ is the
lightest member. In the quark language, $\Theta^+$ is of the
minimal five-quark configuration $\left|uudd\bar{s}\right>$
\cite{qqqqqbar,GM99}, and thus has the exotic strangeness number
S=+1. Predictions about the mass and width of $\Theta^+$ from
chiral soliton models \cite{Pra,penta1,Diak,Weig} have played an
important role in the searches of $\Theta^+$. However, there are
controversies about the parity of $\Theta^+$. Naively, the lowest
$qqqq\bar{q}$ state should have negative parity in quark models
\cite{qqqqqbar,QMDS}. Analysis from QCD sum rule \cite{QCDSR} as
well as earlier lattice QCD calculation \cite{LatticeQCD} show
that $\Theta^+$ can be consistently identified as a pentaquark
state provided its $J^P=\frac{1}{2}^-$. However, analysis of the
stability of pentaquarks $uudd\bar{s}$ shows that \cite{Stan}
$p$-shell with positive parity is lower than $s$-shell with
negative parity for pentaquarks. The quark cluster models
\cite{DQM} also predict a positive parity for $\Theta^+$. In
literatures \cite{Diak,Weig}, the anti-decuplet baryons are
predicted to have positive parity from chiral soliton models.
However, there are also opinions \cite{anti-sol} that the success
predictions from chiral soliton models could be fortuitous. Up to
now, the parity of $\Theta^+$ has not been experimentally
determined. Combining the observed pentaquark $\Theta^+$ mass from
available experiments \cite{LEPS,DIAN,CLAS,SAPH,HERMES} and the
pentaquark $\Xi_{3/2}$ mass from a recent experiment \cite{NA49},
Diakonov and Petrov \cite{Diak2} re-identified N$_{\bar{10}}$ and
$\Sigma_{\bar{10}}$. However, there are no such two corresponding
baryons with $J^P=\frac{1}{2}^+$ in available baryon listings
\cite{PDG}. Thus, they suggested that there must be two new
baryons as the missing members of the anti-decuplet. And in
\cite{AAPSW}, the modified PWA analysis also indicates that the
$N(1710)$ is not the appropriate candidate to be a member of the
anti-decuplet, instead, $N(1680)$ or $N(1730)$ with positive
parity was suggested.
\par The main purpose of this note is to discuss the probability
of picturing $\Theta^+$, $\Xi(1862)$, $N(1650)$ and $\Sigma(1750)$
as the members of an anti-decuplet with negative parity
model-independently by pure symmetry consideration. In soliton
picture, there also exist pentaquark states with one heavy
anti-quark ($\bar{Q}qqqq$) and with $J^{P}=\frac{1}{2}^-$ in the
bound state approach \cite{RS,OPM1}. However, in the collective
quantization approach, how to describe both positive and negative
parity baryons is still not solved.
\par
In the $SU(3)$ chiral soliton model, the fundamental object is the
chiral field $U(x)$
\begin{equation}
U(x)=\exp{\left[i\frac{\lambda_b\phi_b(x)}{f_\pi}\right]},
\end{equation}
where $f_\pi\approx$93~MeV is the observed pion decay constant,
$\lambda_b$ are the eight Gell-Mann $SU(3)$ matrices and
$\phi_b(x)$ are the eight pseudoscalar meson fields. Under the
space inversion transformation, $\phi_b(x)$ transforms as
\begin{equation}
\widehat{P}\phi_b(\mathbf{x},t)\widehat{P}^\dagger=-\phi_b(-\mathbf{x},t),
\end{equation}
where $\widehat{P}$ is the parity operator. Accordingly, $U(x)$
transforms as
\begin{equation}
\widehat{P}U(x)\widehat{P}^\dagger
=\exp{\left[-i\frac{\lambda_b\phi_b(-\mathbf{x},t)}{f_\pi}\right]}=U^\dagger(-\mathbf{x},t)\label{pari}.
\end{equation}
In chiral limit, the action of Skyrme model is of the form
\begin{eqnarray}\nonumber
& I=\frac{f^2_\pi}{4}\int d^4x\mbox{Tr}(\partial_\mu U\partial_\mu
U^\dagger)
\\ &
 +\frac{1}{32e^2}\int d^4x\mbox{Tr}\left[\partial_\mu
UU^\dagger,\partial_\nu UU^\dagger\right]^2+N_c\Gamma,
\end{eqnarray}
which has a hedgehog solitonic solution under the assumption of
maximal symmetry \cite{Skyr,Guad}
\begin{equation}
    U_1(\mathbf{x})=\left(
        \begin{array}{cc}
            \exp{[i(\mathbf{\widehat{r}}\cdot\mathbf{\tau})F(r)]} & \begin{array}{c}0\\0\end{array}\\
            \begin{array}{cc}0&0\end{array}&1
        \end{array}
        \right),
\end{equation}
where $e$ is introduced to stabilize the solitons by Skyrme;
$\Gamma$ is the Wess-Zumino term; $F(r)$ is the
spherical-symmetric profile of the soliton, the solution of the
nonlinear equation of motion; $\mathbf{\tau}$ are the three Pauli
matrices; and $\mathbf{\widehat{r}}$ is the unit vector in space.
The action is invariant under $SU(3)_L\times SU(3)_R$
transformation. However, we are only interested in those $U(x)$
with the same vacuum at $r\rightarrow\infty$
\begin{equation}
    U(x)=A(t)U_1(\mathbf{x})A(t)^{-1}, \ \ A \in SU(3)\label{usky}.
\end{equation}
Inserting (\ref{usky}) into (\ref{pari}) gives
\begin{equation}
\widehat{P}U(x)\widehat{P}^\dagger=U(x),
\end{equation}
and
\begin{eqnarray}
&\widehat{P}A(t)\widehat{P}^\dagger=A(t)\label{apar},\\
&\widehat{P}F(r)\widehat{P}^\dagger=-F(r)\label{fpar}.
\end{eqnarray}
Then quantize the system about this solitonic solution for
collective coordinates $A$. To leading order only symmetry modes
(collective coordinates) are important, thus we only treat
collective coordinates quantum mechanically, and the SU(3)
symmetric effective action in the large $N_c$ limit leads to the
collective Hamiltonian \cite{Guad}:
\begin{equation}
    \widehat{H}=M_{cl}+\frac{1}{2I_2}\left[\widehat{C}^{(2)}-\frac{1}{12}(N_cB)^2\right]+\left(\frac{1}{2I_1}-\frac{1}{2I_2}\right)\mathbf{\widehat{J}}^2,~~
    \label{H}
\end{equation}
where $M_{cl}$, $I_1$ and $I_2$ are given by the
3-dimensional space coordinate integrals of even functions of
$F(r)$ and $e$, and are treated model-independently and fixed by
experimental data in our work. $M_{cl}$ is the classical soliton
mass; $I_1$ and $I_2$ are moments of inertia;
$\widehat{C}^{(2)}$=$\sum\limits_{a=1}^{8}\widehat{G}_a^2$ is the
quadratic (Casimir) operator of the vectorial group SU(3)$_v$, and
in the representation $(p,q)$, its eigenvalue
$C^{(2)}=\frac{1}{3}[p^2+q^2+pq+3(p+q)]$;
$\widehat{G}_a(A)~~(a=1-8)$ are the generators of SU(3)$_v$; and
$\widehat{J}_i(A)~~(i=1-3)$ are the the generators of the spin
group SU(2)$_s$. Using Eq.~(\ref{apar}), we have
\begin{eqnarray}
&\widehat{P}\widehat{H}(A)\widehat{P}^\dagger=\widehat{H}(A),\\
&\widehat{P}\widehat{G}_a(A)\widehat{P}^\dagger=\widehat{G}_a(A),
\ \
\widehat{P}\widehat{J}(A)\widehat{P}^\dagger=\widehat{J}(A)\label{gjpa}.
\end{eqnarray}
The wave function $\Psi_{\nu\nu^\prime}^{(\mu)}$ of baryon $B$ in
the collective coordinates is of the form
\begin{equation}
    \Psi_{\nu\nu^\prime}^{(\mu)}(A)=\sqrt{\mbox{dim}(\mu)}D^{(\mu)}_{\nu\nu^\prime}(A)\label{psi},
\end{equation}
where $(\mu)$ denotes an irreducible representation of the SU(3)
group; $\nu$ and $\nu^{\prime}$ denote $(Y, I, I_3)$ and $(1, J,
-J_3)$ quantum numbers collectively; $Y$ is the hypercharge of
$B$; $I$ and $I_3$ are the isospin and its third component of $B$
respectively; $J_3$ is the third component of spin $J$; and
$D^{(\mu)}_{\nu\nu^\prime}(A)$ are representation matrices.
Eq.~(\ref{apar}) means that the wave function (\ref{psi}) has
positive parity. However, in the procedure of collective
coordinate quantization, only collective parts are quantized, and
modes orthogonal to the symmetry modes are treated classically. In
our case, these modes should be related with the coordinate $r$.
However, by Eq.~(\ref{fpar}), we know that a function of $r$ may
still possess negative parity, this may be suggestive to extend
soliton picture to negative parity baryons because the part we
treat classically may still contribute negative parity. The
symmetry breaking Hamiltonian \cite{Bolt} is
\begin{equation}
    H^\prime=\alpha D^{(8)}_{88}+\beta
    Y+\frac{\gamma}{\sqrt{3}}\sum_{i=1}\limits^3D^{(8)}_{8i}J^i\label{Hp},
\end{equation}
where the coefficients $\alpha$, $\beta$, $\gamma$ are
proportional to the strange quark mass and model dependent, and
are treated model-independently and fixed by experiments; and
$D^{(8)}_{ma}(A )$ is the adjoint representation of the $SU(3)$
group and defined as
\begin{equation}
D^{(8)}_{ma}(A
)=\frac{1}{2}\mbox{Tr}(A^{\dagger}\lambda^mA\lambda^a).
\end{equation}
We can use perturbation theory to calculate the baryon states in
collective coordinates on the basis of flavor symmetry states
(\ref{psi}) \cite{Park}
\begin{eqnarray}
        &&\left|N\right>=\left|N;8\right>+C_{\overline{10}}
        \left|N;\overline{10}\right>+C_{27}\left|N;27\right>,\\
        &&\left|\Sigma\right>=\left|\Sigma;8\right>+C_{\overline{10}}\left|\Sigma;\overline{10}\right>
        +\frac{\sqrt{6}}{3}C_{27}\left|\Sigma;27\right>,\\
        &&\left|\Xi\right>=\left|\Xi;8\right>+C_{27}\left|\Xi;27\right>,\\
        &&\left|\Lambda\right>=\left|\Lambda;8\right>+\frac{\sqrt{6}}{2}C_{27}\left|\Lambda;27\right>,\\
        &&\left|\Theta^{+}\right>=\left|\Theta^{+};\overline{10}\right>,\label{theta}\\
        &&\left|N_{\overline{10}}\right>=\left|N;\overline{10}\right>-C_{\overline{10}}
        \left|N;8\right>+\frac{\sqrt{30}}{80}C_{27}^{(\overline{10})}\left|N;27\right>, ~~~~~~\\
        &&\left|\Sigma_{\overline{10}}\right>=\left|\Sigma;\overline{10}\right>-C_{\overline{10}}
        \left|\Sigma;8\right>+\frac{1}{4\sqrt{5}}C_{27}^{(\overline{10})}\left|\Sigma;27\right>, ~~~~\\
        &&\left|\Xi_{3/2}\right>=\left|\Xi_{3/2};\overline{10}\right>+
        \frac{\sqrt{6}}{16}C_{27}^{(\overline{10})}\left|\Xi_{3/2};27\right>,\label{xi}
\end{eqnarray}
and the coefficients for other multiplets are
\begin{eqnarray}
    &C_{\overline{10}}=-\frac{1}{3\sqrt{5}}(\alpha+
    \frac{\gamma}{2})I_2, \\ &C_{27}=-\frac{\sqrt{6}}{25}(\alpha-
    \frac{\gamma}{6})I_2, \\ &C_{27}^{(\overline{10})}=-(\alpha-\frac{7}{6}\gamma)I_2,
\end{eqnarray}
where $\left|B; \mu\right>$ denotes a flavor symmetry state
$\Psi_{\nu\nu^\prime}^{(\mu)}(A)$ with $\nu\nu^\prime$ denoting
the quantum numbers of baryon $B$. From the analysis above, we see
that this mixing is due to $H^\prime$, and is irrelevant to
parity. Thus  this mixing is still valid if the anti-decuplet
baryons have negative parity. We know that in a baryon multiplet,
different baryon functions are related by
$\widehat{T}_\pm~(=\widehat{G_1}\pm i\widehat{G}_2)$,
$\widehat{V}_\pm~(=\widehat{G_4}\mp i\widehat{G}_5)$ and
$\widehat{U}_\pm~(=\widehat{G_6}\pm i\widehat{G}_7)$, which
commute with $\widehat{P}$ by (\ref{gjpa}). Therefore, to fix the
parity, we can only find a candidate with definite parity from
experiments, which has the same mass and width to the
corresponding member of the anti-decuplet, and then fix the parity
of other baryons in the multiplet accordingly.

Ref.~\cite{NA49} first reported evidence for the existence of a
narrow $\Xi^-\pi^-$ baryon resonance with mass of
$1.862\pm0.003$~GeV and width below the detector resolution of
about $0.018$~GeV, and this state is considered as a candidate for
the pentaquark $\Xi^{--}_{\frac{3}{2}}$ in the anti-decuplet
predicted from chiral soliton models \cite{Diak}. Provided
$\Theta^+$ with mass of 1.54~GeV and $\Xi_{\frac{3}{2}}$ with mass
of 1.86~GeV, we recalculate the coefficients in (\ref{H}) and
(\ref{Hp}) as well as the masses of the other two members of the
anti-decuplet. The results are as follows
\begin{equation}
\begin{array}{lll}
1/I_1=154~MeV;& 1/I_2=399~\mbox{MeV};&\Sigma\approx 78~MeV;\\
\alpha=-663~\mbox{MeV};& \beta=-12~\mbox{MeV};&
\gamma=185~\mbox{MeV};\\
m_{N_{\overline{10}}}=1.65~\mbox{GeV};
&m_{\Sigma_{\overline{10}}}=1.75~\mbox{GeV}\label{data},
\end{array}
\end{equation}
where $\Sigma$ is the pion-nucleon Sigma term. From available
baryon listings \cite{PDG}, there is only one particle, $N(1650)$
with negative parity, for the resonances of $N$ with mass around
1.65~GeV and spin 1/2, and there exists $\Sigma(1750)$ with
$J^P=\frac{1}{2}^-$ for the resonances of $\Sigma$ with mass
around 1.75~GeV. It has been also found \cite{wu_ma1_2}
surprisingly that all the nonexotic members of the 27-plet with
spin 3/2, calculated with the above parameters, can be identified
from the available baryon listings~\cite{PDG}, agreeable with
experiments both in mass and width.

Inspired by the above observation, we want to know whether there
exists some symmetry between those four baryons of $\Theta^+$,
 $N(1650)$, $\Sigma(1750)$, and $\Xi(1862)$, assuming them
as a set of members in an anti-decuplet. Since we still do not
know how to investigate it from chiral soliton picture, we want to
check it by symmetry consideration model-independently. For the
decay of $J^-\rightarrow J^{+}+0^-$, we have a following effective
interaction langrangian

\begin{equation}
\mathcal{L}_I=ig\bar{\psi}\psi\varphi_m,
\end{equation}
and from it we have
\begin{widetext}
\begin{eqnarray}
    \Gamma(B\rightarrow B^\prime m)=\frac{G_s^2}{4\pi}\frac{m_B^\prime}{m_B}|\mathbf{p}|\left\{\begin{array}{c}
    \frac{\mbox{dim}(\mu^\prime)}{\mbox{dim}(\mu)}\left|\begin{array}{c}\sum\limits_\gamma\left(
    \begin{array}{cc}8&\mu^\prime\\ Y_mI_m&Y_\rho I_\rho \end{array}\right|
    \left.\begin{array}{c}\mu_\gamma\\Y_\nu
    I_\nu\end{array}\right)
    \left(
    \begin{array}{cc}8&\mu^\prime\\00&1 J_\rho \end{array}\right|
    \left.\begin{array}{c}\mu_\gamma\\1J_\nu\end{array}\right)\end{array}\right|^2\end{array}\right\},\\\label{w2}
\end{eqnarray}
\end{widetext}
where $\varphi_m$ is any pseudoscalar meson, $\bar{\psi}$ is a
baryons belonging to baryon multiplet $(\mu)$ with negtive parity,
and $\psi$ is a baryons belonging to baryon multiplet
$(\mu^\prime)$ with positive parity. In this paper, we will only
deal with an anti-decuplet with negative parity. Thus we get the
widths of anti-decuplet baryons in Table 1 by using this formula
and identifying $\Theta^+$, $N(1650)$, $\Sigma(1750)$, and
$\Xi(1862)$ as the members of the anti-decuplet, and the results
calculated in the case of positive parity from chiral soliton
models are still list in Table~1.
\begin{widetext}
\vspace{1cm} \begin{center}
\begin{tabular}{llll}
\multicolumn{4}{c}{Table~1. The widths of baryons in the
anti-decuplet}\\\hline
Mode&estimation(MeV)&$J^P=\frac{1}{2}^-$(MeV)&$J^P=\frac{1}{2}^+$(MeV)\\\hline
$\Theta^+\rightarrow
KN$&$<25$&25$(\mbox{input})$&25$(\mbox{input})$\\\hline
N(1650)$\rightarrow N\pi$&$80\sim$171&12&25\\
N(1650)$\rightarrow N\eta$&$4\sim19$&8&8\\
N(1650)$\rightarrow\Lambda K$&$4\sim21$&4&1.5\\\hline
$\Sigma(1750)\rightarrow N\bar{K}$&6$\sim$64&7&3\\
$\Sigma(1750)\rightarrow \Sigma\pi$&$<$12.8&8&14\\
$\Sigma(1750)\rightarrow\Sigma\eta$&9$\sim$88&2&0.2\\
$\Sigma(1750)\rightarrow\Lambda\pi$&seen&12&28\\\hline
$\Xi_{3/2}\rightarrow\Xi\pi$&$<$18(?)&24&46\\
$\Xi_{3/2}\rightarrow\Sigma\bar{K}$&&18&15\\\hline
\end{tabular}
\end{center}
\end{widetext}

In summary, we recalculated the masses and widths of the baryons
in the anti-decuplet. The calculated results show that if we
accept $\Theta^+$ and the reported particle $\Xi(1862)$ in
\cite{NA49} as members of the anti-decuplet, we will get the mass
of $N_{\bar{10}}$ around 1.65~GeV and the mass of
$\Sigma_{\bar{10}}$ around 1.75~GeV. From available baryon
listings \cite{PDG}, we find that there are only $N(1650)$ and
$\Sigma(1750)$ with negative parity. And inspired by this, we give
a width formula for the negative parity baryon decay on symmetry
consideration and give all the widths of baryons of an
anti-decuplet with negative parity but the same masses as
calculated from chiral soliton models (Table~1). From Table~1, The
width of $\Sigma(1750)$ in the case of $J^P=\frac{1}{2}^-$ fits
experimental data better, but for $N(1650)$, there is an obvious
deviation in the process $N(1650)\rightarrow N\pi$ while the
widths agree with other two decay channels $N(1650)\rightarrow
N\eta$ and $N(1650) \rightarrow\Lambda K$. There could be a
possibility for a missing negative parity $N_{\bar{10}}$ with mass
around 1650~\mbox{MeV} and a narrow width or that the SU(3)
breaking plays a role. The width of $\Xi_{3/2}$ with negative
parity also agrees better with experimental observation. In the
quark picture of the anti-decuplet, $N_{\bar{10}}$ and
$\Sigma_{\bar{10}}$ contain hidden quark-antiquark pairs, while
$N(1650)$ and $\Sigma(1750)$ are the orbital angular excitations
in the quark model. Thus, to determine the parity of the
anti-decuplet, we need further experiments to find missing
non-exotic members and also to determine their parities.

We are grateful for helpful discussions with Yajun Mao, Zhi-Yong
Zhao, Han-Qing Zheng, and Shi-Lin Zhu. This work is partially
supported by National Natural Science Foundation of China under
Grant Numbers 10025523 and 90103007.


\end{document}